\documentclass[amsmath,amssymb,aps,prd,onecolumn,showpacs,showkeys,preprint]{revtex4}
\usepackage{hyperref}

\begin{document}
\title{
 Measuring gravitational behavior  at short distances in space: \\A local test for MOND/MOG
}

\author{Qasem Exirifard}
\email{exir@theory.ipm.ac.ir}

\begin{abstract} 
We consider the AQUAL theory - a theory of modified gravity capable of resolving the missing mass problem - and study its predictions for micro gravity tests at the gravitational saddle points of the Solar system.  We report that the AQUAL model enhances the gravity at the sub-micrometer ranges around the gravitational saddle points in a way that so far has been unnoticed.  This enhancement can be measured. We, therefore, call for moving toward  implementing micrometer gravity tests within the Solar gravitational saddle points.
\end{abstract}

\pacs{04.50.Kd, 04.80.Cc, 95.35.+d}
\keywords{Theories of modified gravity, MOND, gravity at small distances, test of gravity, dark matter}
 \preprint{arXiv:1206.0173} 
\maketitle
Theories of modified gravity are one of the possible candidates for the resolution of  the missing mass problem in galaxies \cite{Famaey:2011kh}. One of the most successful  of these theories is the AQUAL model \cite{AQUAL} which assumes that gravity is not governed by the Newton's law
\begin{equation}
\Box \phi \neq 4 \pi G \rho\,,
\end{equation}
where $\phi$ is the gravitational potential and $\rho$ is the matter's density. It assumes that gravity is governed by 
\begin{equation}\label{AE}
\nabla_i~\left( \mu(\frac{|\nabla\phi|}{a_0})\nabla^i \phi\right) = 4 \pi G \rho\,,
\end{equation}
wherein  the functional $\mu$ is demanded to  possess the following asymptotic behaviors
\begin{eqnarray}\label{mu}
\mu(x) = \left\{
\begin{array}{ccc}
1&,& x\gg 1 \\
x &,& x \leq 1
\end{array}
\right.\,,
\end{eqnarray} 
and $a_0$ reads
\begin{equation}\label{a0}
a_0 = (1.0\pm0.2)  \times 10^{-10} \frac{m}{s^2}\,.
\end{equation}
The AQUAL model is constructed to provide a platform for a naturally relativistic realization of  the MOND paradigm \cite{MOND}. It is  also free of some of the conceptual problems of the MOND. 
The solutions to the AQUAL gravity can be written in term of the solutions of the Newtonian gravity:  
 \begin{eqnarray}\label{EMond}
\mu(\frac{|\nabla\Phi|}{a_0}) \nabla\Phi  & = & \nabla\Phi_{N} + \nabla\times \vec{{h}} \,,
\end{eqnarray}
where $\vec{{h}}$ is identified  by
\begin{eqnarray}\label{eqhe}
0\,=\,\nabla \times \nabla\Phi  & = & \nabla\times(\frac{ \nabla\Phi_{N} + \nabla\times \vec{{h}}}{\mu(\frac{|\nabla\Phi|}{a_0})})\,.
\end{eqnarray}
It is expected that $|\nabla\times h|< |\nabla \phi_N|$ in all general physical cases. When large amount of symmetry is present as for the spherical symmetry then $\vec{h}$ vanishes. We also note that  $\vec{h}$ vanishes  when $\nabla \phi_N$ is constant.   

 It is known that the AQUAL theory can be tested within the Solar system. This is because the Solar system is a composed system: as the Sun follows its orbit around  the center of the Milky way and the planets revolve  around the Sun in some places within the Solar system the Newtonian gravitational field vanishes. Let the surrounding of  these places wherein the AQUAL theory governs the dynamics of gravity  (wherein $\mu(x)\approx x$) be called the MOND windows. It  has been suggested that an accurate tracking of a probe like the LISA path finder that passes through the MOND windows can prove or refute the AQUAL theory \cite{Bekenstein:2006fi,Mozaffari:2011ux,Magueijo:2011an}. 

Ref \cite{Bekenstein:2006fi} within its section IX briefly considers the possibility of measuring the behavior of the gravity on board within the MOND windows.  In this letter we reconsider performing an on-board experiment within the MOND windows in details.  In addition to the known result, we report a  new prediction for the AQUAL model. We report an enhancement of the effective Newtonian constant measured by appropriate on-board experiments. 

The on-board experiment must be implemented by an apparatus which is producing a time-dependent gravitational acceleration of a definite frequency. We present the time-dependent gravitational field by $\phi_{lab}$. We present the rest of the gravitational field by $\phi_{ext}$ and we call it the external gravitational field. In so doing we decompose the gravitational field to $\phi\equiv \phi_{ext}+\phi_{lab}$. Inserting the decomposition into \eqref{AE} holds:
\begin{equation}\label{final}
\nabla_i~\left( \mu(\frac{|\nabla\phi_{ext}+\nabla\phi_{lab}|}{a_0})\nabla^i (\phi_{ext}+\phi_{lab})\right) = 4 \pi G (\rho_{ext}+\rho_{lab})\,,
\end{equation}
wherein $\rho_{lab}$ is the fluctuating density producing the laboratory's gravitational field of the definite frequency and $\rho_{ext}$ is the rest of the mass density.  We consider solving the above equation for the cases where  
\begin{eqnarray}
|\nabla\phi_{ext}|&<& a_0\label{PM}\,,\\
|\nabla\phi_{lab}|&<& |\nabla\phi_{ext}| \label{EL}\,.
\end{eqnarray}
Performing the experiment within the Solar MOND windows holds \eqref{PM}.  \eqref{EL} requires measuring a very weak gravitational force. The very low non-gravitational acceleration  $10^{-11} \frac{m}{s^2}$ \cite{MONDtestEarth1} and $10^{-14}\frac{m}{s^2}$ \cite{MONDtestEarth2} have been experimentally measured. The gravitational acceleration of $10^{-10}\frac{m}{s^2}$ has been measured \cite{Meyer:2012ae}. We assume that in future a gravitational acceleration about or less than   $(10^{-12}-10^{-11}) \frac{m}{s^2}$ can be  experimentally  measured.  We also consider that within the volume wherein the apparatus performs the measurement,  the external gravitational field can be approximated to a constant field:
\begin{equation}\label{constant}
\nabla \phi_{ext}\approx cte\,.
\end{equation} 
When we are measuring the gravitational field at the micrometer ranges we can approximate the external gravitational field to a constant field. This justifies \eqref{constant} .  Thanks to the theory of large extra dimensions \cite{ArkaniHamed:1998rs}, we have seen technological advances in measuring the behavior of gravity at the micrometer ranges, such as \cite{g,g0,g1}. We assume that these technologies can be advanced in order to measure a gravitational acceleration of order or less than    $10^{-12} \frac{m}{s^2}$ in the space.

Utilizing \eqref{EL} in the  taylor expansion of $\mu$ in  term of $\frac{|\nabla \phi_{ext}|}{|\nabla\phi_{lab}|}$  yields
\begin{equation}\label{MuExpand}
\mu(\frac{|\nabla\phi_{ext}+\nabla\phi_{lab}|}{a_0}) = \mu(\frac{|\nabla\phi_{ext}|}{a_0}) +  \frac{\mu'(\frac{|\nabla\phi_{ext}|}{a_0})}{a_0 |\nabla\phi_{ext}|} \nabla \phi_{ext}.\nabla\phi_{lab} + O(|\nabla\phi_{lab}|^2)\,.
\end{equation}
wherein $\mu'(x)= \frac{d\mu}{dx}$. Inserting \eqref{MuExpand} into \eqref{final} and separating the leading and the sub-leading terms gives: 
\begin{eqnarray}\label{exteq}
\nabla_i~\left( \mu(\frac{|\nabla\phi_{ext}|}{a_0})\nabla^i \phi_{ext}\right) &=& 4 \pi G \rho_{ext}\,,\\ 
\label{labeq}
\nabla_i\left(\mu(\frac{|\nabla\phi_{ext}|}{a_0})\nabla^i \phi_{lab}\right) + 
\nabla_i\left(\frac{\mu'(\frac{|\nabla\phi_{ext}|}{a_0})}{a_0|\nabla \phi_{ext}|}\nabla \phi_{ext}.\nabla\phi_{lab} \nabla^i\phi_{ext}\right)&=&  4 \pi G \rho_{lab}\,.
\end{eqnarray}
Notice that \eqref{exteq} is the equation of motion for the external field while \eqref{labeq} is the equation of the motion for the laboratory's field.  Utilizing \eqref{constant} then simplifies \eqref{labeq} to 
\begin{eqnarray} \label{eqlab}
\mu(\frac{|\nabla\phi_{ext}|}{a_0}) \Box \phi_{lab} +  \frac{\mu'(\frac{|\nabla\phi_{ext}|}{a_0})}{a_0 |\nabla\phi_{ext}|}\nabla^i \phi_{ext}\nabla^j \phi_{ext}\nabla_i\nabla_j\phi_{lab} &=& 4 \pi G \rho_{lab}\,.
\end{eqnarray}  
wherein all the quantities but $\phi_{lab}$ and $\rho_{lab}$ are now constant. So \eqref{eqlab} can be expressed in the following form 
\begin{equation}\label{NOS}%NonOrthogonalSystem
g_{ij} \partial^i \partial^j \phi_{lab} =  4 \pi \, G_{{eff}}\, \rho_{lab}\,,
\end{equation}
wherein 
\begin{eqnarray}
g_{ij} & \equiv & \eta_{ij} + h_{ij}\,, \\
h_{ij} & \equiv & \frac{\mu'(\frac{|\nabla\phi_{ext}|}{a_0})}{\mu(\frac{|\nabla\phi_{ext}|}{a_0})}
\frac{\nabla^i \phi_{ext}\nabla^j \phi_{ext}}{a_0|\nabla \phi_{ext}|} \,,
\end{eqnarray}
and 
\begin{equation}\label{Geff}
G_{eff}= \frac{1}{\mu(\frac{|\nabla\phi_{ext}|}{a_0})}\, G \,.
\end{equation}
Eq. \eqref{NOS} is the Laplace equation written in a non-orthogonal coordinate system. The eigenvalues of the metric $g$  follows:
\begin{equation}
\text{Eigenvalues}[g_{ij}]=[1,1,1+ \frac{\mu'(\frac{|\nabla\phi_{ext}|}{a_0})}{\mu(\frac{|\nabla\phi_{ext}|}{a_0})} \frac{|\nabla \phi_{ext}|}{a_0}]\,.
\end{equation}
In the Newtonian regime where $\mu(x)=1$, the eigenvalues become trivial. Deviation from one in one of the eigenvalues is a facet of the known external field effect. In the deep MONDian regime wherein $\mu(x)=x$, the eigenvalues changes to $(1,1,2)$. This factor of two also has been noticed in \cite{Bekenstein:2006fi}.  Let it be highlighted that perhaps the interpolating function $\mu$ must be chosen such that $1+\frac{\mu'(x)\, x}{\mu(x)} $ remains positive, finite and non-zero for all positive $x$.

Now let us choose  the eigenvectors of $g_{ij}$ as  the new coordinate bases. In this new coordinate then  \eqref{NOS} converts to the more familiar form
\begin{equation}\label{NOS1}%NonOrthogonalSystem
\tilde{\Box} \phi_{lab} =  4 \pi \, G_{{eff}}\, \rho_{lab}\,,
\end{equation}
where $\tilde\Box$ is the ordinary Laplace operator in the semi-Cartesian presentation.  \eqref{NOS1}  clearly states that the gravitational constant is altered.  The effective gravitational constant \eqref{Geff} in the deep MOND regimes due to \eqref{mu} simplifies to   
\begin{equation}
G_{eff}= \frac{a_0}{|\nabla\phi_{ext}|}\, G \,.
\end{equation}
The above relation states that the AQUAL model in the considered setup replaces  the Newton constant by an effective constant. In fact, due to \eqref{PM}, it predicts an enhancement for the Newton's constant. This Letter  reports this enhancement for the first time. The enhancement is more for weaker external gravitational acceleration. %Note that the enhancement is not an artifact of the coordinate transformation employed to simplify the equations.  We first fix the coordinate on the earth where the gravitational field is strong. Then the external gravitational field dictates the coordinate wherein the equations are simplified to \eqref{NOS1} in the MOND windows. This fact can also be directly deduced from \eqref{eqlab}: just imagine an  experimental apparatus  producing a $\nabla\phi_{lab}$  perpendicular to the $\nabla\phi_{ext}$. 

The largest MOND window occurs around  the gravitational saddle point of the Sun and Jupiter where the gravitational field of the Sun and Jupiter cancel each other.  The size of the Sun-Jupiter MOND window is about $5 km \times (140 km)^2$. This window is large to accommodate a satellite or a probe \footnote{Ref. \cite{Exirifard:2011bz} proposes that measuring  the gravitomagnetic force in the gravitational saddle points of the Solar system proves or refutes modified theories of gravity.}.  We propose and call for measuring the behavior of gravity at the micrometer ranges at the neighborhood of the gravitational saddle point of the Sun-Jupiter system, or any other gravitational saddle point.  For example   provided that 
\begin{enumerate}
\item A probe is sent to the gravitational saddle point of the Sun-Jupiter system and kept for sufficient amount of time  in a bubble wherein the total external gravitational field is less than $10^{-11} \frac{m}{s^2}$. Note  that the approximate size of this bubble is $500 m * (10km)^2$. This bubble leads to an enhancement of factor ten for the gravitational constant.
\item Micrometer gravity reaches the sensitivity of measuring a gravitational acceleration of $10^{-12} \frac{m}{s^2}$ and below. This low acceleration should suffice to drop the sub-sub-leading corrections in the expansion of \eqref{final} in term of $\frac{|\nabla \phi_{lab}|}{|\nabla \phi_{ext}|}$ with a confidence level about 10 sigmas.  
\item The Newton's constant can be measured at least with the error bar of 10 percent through analyzing the behavior of gravity at the micrometer ranges.
\end{enumerate}
 Then  measuring the behavior of  the gravity at the micrometer ranges within the gravitational saddle point of the Sun-Jupiter system proves or refutes the approach of modified theories of gravity to the missing mass problem in galaxies at a confidence level beyond 10 sigmas. This suggests that all the future probes that will pass through the MOND solar windows be equipped with appropriate modules capable of measuring the behavior of gravity at the micrometer ranges. It is also  very interesting to add these modules to any future probes planned to be sent outside the solar system in order to gain experimental/empirical data on the behavior of gravity in the limit  of a very weak external field. 
 
  \section*{Acknowledgements}
 This work was supported by the Institute for Research in Fundamental Sciences.

\providecommand{\href}[2]{#2}\begingroup\raggedright

\end{document}